\documentstyle[prb,preprint,aps,epsf]{revtex}
\tightenlines
\begin{document}
%
%
%
\title{Reciprocal Space Analysis of Short Range Order Intensities by the
Cluster
Variation Method}
\author{J.M. Sanchez}
\address{Center for Material Science and Engineering\\
The University of Texas at Austin, Austin, Texas 78712}
\author{V. Pierron-Bohnes}
\address{Institut de Physique et Chimie des Mat\'eriaux de Strasbourg\\
UMR46 CNRS-Universit\'e Louis Pasteur-EHICS 23, 67037
Strasbourg, France}
\author{F. Mej\'{\i}a-Lira}
\address{Instituto de F\'{\i}sica\\
Universidad Aut\'onoma de San Luis Potos\'{\i}, 78000 San Luis Potos\'{\i},
S.L.P., Mexico}
\date{\today}
\maketitle
%
\begin{abstract}
A reciprocal space formulation of the Cluster Variation is used in order to
extract effective pair interactions in alloys from experimental short-range
order diffuse intensities. The method is applied to the analysis of the
short range order contribution to the neutron diffuse scattering of a
$Fe-19.5 {\%} Al$ single crystal. A detailed comparison with real space
methods is carried out for three different levels of approximations. For
the highest level of approximation used in this study, effective pair
interactions up to fifth neighbors are obtained.
\end{abstract}
%
%
%
\newpage
%
%

\section{Introduction}
The diffuse scattering of X-rays and neutrons remains the technique of
choice for quantitative studies of short-range order (SRO) in alloys. In
particular, the part of the diffuse intensity related to substitutional SRO
has played a prominent role in alloy theory. This SRO intensity contains
important information on the effective interactions between atoms which, in
turn, can be used to characterize thermodynamic properties of the alloy,
such as relative phase stability, phase diagrams and antiphase boundary
energies.

Experimental measurements of SRO intensity have evolved significantly from
the studies in polycrystalline materials of the early 1950's. At present,
accurate measurements are carried out in single crystals at high and low
temperature. For example, recent work has focused on the use of high flux
neutron and tunable synchrotron X-ray sources.
\cite{bess86,butl89,sola89,pier90,rein90,pier91,rein92} On the theoretical
side, the study of SRO intensity in alloys and its characterization through
diffuse scattering dates back to the pioneering work of Cowley,
\cite{cowl50} Krivoglaz \cite{kriv69} and Clapp and Moss.
\cite{clap66,moss68} Out of these studies emerged what is presently known
as the Krivoglaz-Clapp-Moss (KCM) formula. This formula establishes a
simple relationship between the experimentally observed diffuse scattering
and the Fourier transform of the effective pair interactions. The KCM
formula corresponds to a treatment of concentration fluctuations at the
lowest level of approximation for the alloy configurational free energy,
namely the single site mean-field approximation, also known as the
Bragg-Williams \cite{brag34} and the molecular field approximation.

More recently, effective pair interactions have been obtained from
experimentally determined SRO intensities using higher approximations based
on Monte Carlo simulations and the Cluster Variation Method (CVM).
\cite{kiku51} The first calculation of SRO intensity carried out directly
in reciprocal space using the cluster variation method was done by Sanchez
in 1982. \cite{sanc82} This k-space formulation of the CVM provides a
significant improvement over the KCM formula, although a price is paid in
increased computational effort. Thus, despite the improved treatment of
fluctuations by the CVM, the method has been applied only to very few
cases. \cite{sanc82,mohr85} An alternative approach, which we will refer to
as the real space inverse CVM was introduced by Gratias and C{\' e}n{\'
e}d{\` e}se. \cite{grat85} In this method, the effective pair interactions
are obtained by fitting a few Warren-Cowley SRO parameters which, in turn,
are related by a Fourier transform to the experimentally determined SRO
intensity. A definite advantage of the real space inverse CVM is that the
method involves only a modest computational effort, comparable to that of a
regular CVM equilibrium calculation. On the other hand, the main drawback
of the real space technique is that the fitting is done for only a few
Warren-Cowley SRO parameters which, typically, are not enough to describe
the measured intensities. To a great extent, this problem is solved by
applying the inverse method in real space using Monte Carlo simulations, in
which longer interaction ranges may be included without any fundamental
difficulties. \cite{live87,schw88} A problem of a slightly different nature
is the fact that the determination of the Warren-Cowley SRO parameters from
the experimental intensities, despite the fact that they are uniquely
defined in terms of a Fourier transform, is not without errors. This step
in the procedure, which of necessity requires the introduction of a real
space cut-off in SRO, is an additional source of uncertainty on the derived
effective pair interactions.

Here we proposed a new method that carries out the fitting of the SRO
intensity directly in k-space using the CVM formulation of the SRO
intensity of Sanchez. \cite{sanc82} Since the Warren-Cowley SRO parameters
are not required, the k-space fitting circumvents two of the main problems
of the real space method, namely the cut-off in SRO and the determination
of the SRO parameters which, as mentioned, must be obtained by either
fitting or by Fourier transformation of the experimental data. The method
is applied to diffuse neutron scattering data obtained at several
temperatures for a $Fe-19.5at {\%} Al$ single crystal.\cite{pier91} A
comparison between the real and k-space methods is also carried out for
three different approximations of the CVM.

The organization of the paper is as follows. In Section II we describe
briefly the experimental aspects. The theoretical model and a review of the
CVM formulation of the SRO intensity is given in Section III. Results are
presented in Section IV with concluding remarks given in Section V.

\section{Experimental Considerations}

The experimental results, including standard corrections to the data, have
been reported in previous publications.\cite{pier90,pier91} Thus, here we
briefly recall only the main experimental aspects, referring the reader to
Refs. [\onlinecite{pier90,pier91}] for further details.

The diffuse scattering measurements were carried out at high temperatures
on D7 at ILL (Grenoble, France), using a 10 mm in diameter and 20 mm long
cylindrical single crystal of $Fe-19.5at{\%} Al$ oriented along the [110]
direction. Time of flight analysis was used in order to eliminate the
phonon and critical magnetic contributions. Here we analyze four
temperatures between $1273 K$ and $1573 K$. The cross sections were
calculated using the classical corrections (instrumental background, double
scattering and absorption). The experimental points were averaged with a
weight inversely proportional to the square of their error to obtain a
regular grid of points (mesh of 0.1 reciprocal lattice unit step).

In order to obtain the SRO intensity, the experimentally determined neutron
scattering cross section was corrected for contributions due to static and
dynamic atomic displacements. The correction for static displacements was
carried out using the method based on the formulation of Borie and Sparks,
\cite{bori71} whereas dynamic attenuation was treated using the standard
Debye-Waller theory. It should be noted that the accuracy of these
corrections affect critically the reliability of the effective interactions
derived from the diffuse intensity.

In terms of the experimental scattering cross section, the corrected
intensity in Laue units is written:

\begin{equation}
 {I_{corr}(\vec k)} = {1 \over {N F_{Laue}} } \left[\displaystyle {{d
\sigma} \over {d \Omega}}( \vec k) \enskip e^{(B k^2 / 8 \pi^2)} -
I_{inc}\right]
\label{eq1}
\end{equation}

\noindent where $N$ the total number of atoms and $ F_{Laue}$ is the usual
normalization factor given by $ F_{Laue} = c(1-c)(b_{Al} - b_{Fe})^2 $, with
$c$ the aluminum concentration, and $b_{Al}$ and $b_{Fe}$ the nuclear
diffusion lengths of $Al$ and $Fe$, respectively. In Eqn.\ (\ref {eq1}),
$I_{inc}$ is the incoherent scattering and $B$ the Debye-Waller factor.

Following Borie and Sparks, \cite{bori71} the corrected intensity is
approximated by a first order expansion in the atomic displacements $\vec
u_{\bf p}$, where the subindex ${\bf p}$ stands for the set of integers
$(p_1,p_2,p_3)$ and the location of lattice site ${\bf p}$ is given by
$\vec R_{\bf p} = {\displaystyle {a \over 2}} (p_1,p_2,p_3)$. In this
approximation, the corrected intensity at the reciprocal space vector $
\vec k = \displaystyle {{2 \pi} \over a } (h_1,h_2,h_3) $ can be written
as:

    \begin{equation}
         I_{corr}(\vec k) = \alpha (\vec k) \enskip + \enskip
         \sum_{i=1}^{3} h_i Q_i(\vec k)
 		\label{Icorr}
	\end{equation}

\noindent where $\alpha (\vec k) $ is the SRO contribution we are
interested in, and the $Q_i(\vec k) $ are related to the Fourier transform
of the atomic displacements. We next consider each of these two terms
separately.

The SRO intensity $ \alpha (\vec k) $ is given by the Fourier transform
of the Warren-Cowley SRO parameters $\alpha_{\bf p}$:

	\begin{equation}
        \alpha (\vec k) = \sum_{\bf p} \enskip \alpha_{\bf p} \enskip
          cos (\vec k \cdot \vec R_{\bf p})
		\label{alphak}
	\end{equation}

\noindent with $\alpha_{\bf p}$ defined by:

   \begin{equation}
     \alpha_{\bf p} = {{<\sigma_{\bf o} \sigma_{\bf p}> - < \sigma_{\bf o}>^2}
     \over { 1 - < \sigma_{\bf o}>^2}}
     \label{alphap}
     \end{equation}

\noindent where $\sigma_{\bf o}$ and $\sigma_{\bf p}$ are occupation
numbers at the origin and at site $\bf p$, respectively. These occupation
numbers take values $1$ or $-1$ if the lattice site is occupied by $Fe$ or
$Al$ atoms, respectively, and the brackets $<>$ stand for configurational
averages.

The quantities $\vec Q(\vec k)=[Q_1,Q_2,Q_3]$ in Eqn.\ (\ref {Icorr}) are
given by the first order displacement parameters $\vec \gamma_{\bf p}$:

\begin{equation}
	\vec Q(\vec k) =   \sum_{\bf p}
	\enskip \vec \gamma_{\bf p} \enskip sin(\vec k \cdot \vec R_{\bf p})
	\label{Qi}
\end{equation}

\noindent In turn, the displacement parameters $\vec \gamma_{\bf p}$ are
defined by:

\begin{equation}
     \vec \gamma_{\bf p} = - {{2 \pi} \over a }  \enskip
     \sum_{\sigma = \pm 1} \sum_{\sigma ' = \pm 1}
     \enskip {b_{\sigma} b_{\sigma '} \over F_{Laue}} \enskip
     \rho_{2,{\bf p}}(\sigma \sigma ') \enskip <\Delta
       \vec u_{\bf p}^{\sigma \sigma '}>
	\label{gamma}
\end{equation}

\noindent where $<\Delta \vec u_{\bf p}^{\sigma \sigma '}> $ is the average
relative displacement between atoms of type $\sigma $ and $\sigma '$ (i.e.
$Fe$ or $Al$) separated by $\vec R_{\bf p}$, and where $ \rho_{2,{\bf
p}}(\sigma \sigma ')$ is the probability of finding atoms $\sigma $ and
$\sigma '$ separated by distance $\vec R_{\bf p}$. We note that these
probabilities are $c^2 + c(1-c) \alpha_{\bf p}$, $(1-c)^2 + c(1-c)
\alpha_{\bf p}$ and $c(1-c) - c(1-c) \alpha_{\bf p}$ for $AlAl$, $FeFe$ and
$FeAl$ pairs, respectively.

The quantities $\alpha (\vec k) $ and $\vec Q (\vec k)$ are invariant to
any symmetry operation of the space group of the reciprocal lattice,
whereas the diffuse intensity $I_{corr}(\vec k)$ is not. Consequently, the
experimental points can be grouped into families of equivalent k-points
having the same values of $\alpha (\vec k) $ and $\vec Q (\vec k)$, but
with different intensities $I_{corr}(\vec k)$. Families having as many or
more k-points than unknowns ($\alpha (\vec k) $ and $\vec Q (\vec k)$) can
be used to separate both contributions from $I_{corr}(\vec k)$.

Since for the (110) reciprocal plane $Q_{1}= Q_{2}$, there are at most 3
unknowns on this plane. We note that at special points the number of
unknowns is further reduced by symmetry to either 2, when $Q_1=Q_2=Q_3$, or
1, when $Q_1=Q_2=Q_3=0$. The maximum number of independent measurements
possible at some of the equivalent points is 6. However, the measurement
range in reciprocal space is limited by the neutron wavelength and by the
lack of accuracy in the diffuse scattering around the Bragg peaks due to
the mosaic structure of the single crystal and to different parasitic
contributions. These limitations reduce strongly the occurrence of 6
equivalent measurement points. In fact, in the more frequent case there are
4 experimental points for three unknowns. In such cases (24 measured
points), the unknowns were solved using a weighted least square fit, with a
weight inversely proportional to the square of the experimental error.
Thus, for each such point in the first Brillouin zone, one obtains a value
$\alpha (\vec k)$ with an error bar, $\Delta \alpha (\vec k)$, and a
measure of the quality of the fit, $\chi^2$, given by the weighted sum of
the square of the difference between the measured and fitted values. In
order to take into account the quality of the fit, the error bar $\Delta
\alpha (\vec k)$ was multiplied by $\chi$ when the latter was larger than
one. For 18 k-points, the number of unknowns is equal to the number of
experimental points. In these cases the system was solved and the error bar
was multiplied by the maximum $\chi$ of the first class of points. Finally,
for ten points, mainly located around the origin and at the edges of the
measured region, there are less measurements than unknowns and, therefore,
it is not possible to separate $\alpha (\vec k)$ and $\vec Q (\vec k)$ from
the measured intensity.

A point of clarification concerning the Debye-Waller factors is also in
order. These were measured from the attenuation of the fundamental Bragg
peaks through neutron and gamma ray scattering. \cite{pier89} We note,
however, that the attenuation of the Bragg peaks is related to the sum of
the dynamic, $B_{d}(T)$, and static, $B_{s}$, Debye-Waller factors, i.e.
$B_{lro} = B_{d}(T) + B_{s}$, whereas the attenuation of the short range
order intensity is given by their difference, $B = B_{d}(T) - B_{s}$. In
order to separate these two contributions, the static part was assumed to
be temperature independent. Since this approximation results in large
errors in $B_{d}(T)$ at high temperature, the restriction of a temperature
independent static Debye-Waller factor was subsequently lifted by including
a correction $\Delta B(T)$ in the SRO attenuation: \cite{pier89}

\begin{equation}
	B = B_{d}(T) - B_{s} + \Delta B(T)
	\label{D-W}
\end{equation}

The correction $\Delta B(T)$ was determined by exploiting the symmetry of
the corrected intensity as well as using a least square fit of the
experimental data at equivalent k-points. The latter procedure was carried
using points for which the number of unknowns is smaller than the number of
measurements.

The values of $\Delta B(T)$ obtained by the two different methods are very
similar at high temperatures, indicating that the Borie and Sparks model is
adequate to describe the scattering cross section. Departure are seen at
low temperatures, likely due to the magnetic short range order
contributions which are not explicitly considered in the analysis. Thus,
the high temperature data reported in Refs.[\onlinecite{pier91}] are
considered to be the most accurate, and we will consequently restrict the
present analysis to them.

Figure 1 shows the experimental SRO intensity at $1273 K$ after all
corrections have been made, together with the fit obtained with 8
Warren-Cowley SRO parameters (solid line). In what follows, we will refer
to these quantities as the experimental Warren-Cowley SRO parameters,
although they are actually the result of fitting the $\alpha(k)$. In the
fit, a value $\alpha_0 = 1.147$ was obtained, which is to be compared with
the exact value of 1. Each integer step on the horizontal axis of Fig.\ 1
corresponds to a scan along the $(00h_3)$ direction, with $h_3$ between 0
and 1, for fixed values of $h_1 = h_2$. We note that the SRO data at the
other temperatures analyzed here, namely $1373K$, $1473K$ and $1573K$, look
qualitatively the same. Furthermore, at these temperatures, a similar
quality of the fit is also achieved with 8 shells of $\alpha_n$'s.

A notable feature of the SRO intensity is the small peak observed near the
origin. It is not clear from the experimental data, however, whether this
maximum is real or it is simply an artifact due to the cut-off imposed on
the SRO parameters and/or to the fact that the fitting was carried out on
only one plane in reciprocal space. The presence of this maximum, whether
real or not, poses interesting questions on the use of the real space
inversion methods. We will return to this point after discussing the
theoretical models used to describe the SRO intensity in alloys.

\section{Theory}

In this section we develop a simple statistical model in order to calculate
the observed SRO diffuse intensity in terms of the effective pair
interactions. We begin by describing the Hamiltonian, which is to be solved
at finite temperatures using different levels of approximation of the CVM,
continue with a brief description of the approach used to calculate the SRO
intensity, and conclude with a description of the inverse methods employed
to determine the pair interactions.

\subsection{The Hamiltonian}

We describe the $Fe-Al$ system with a simple Ising model in which the
magnetic moments are localized on the $Fe$ atoms:

\begin{eqnarray}
	H_0 =- \frac{1}{8} \enskip \sum_{{\bf
	p},{\bf p'}} \enskip J_{\bf pp'} \enskip (1 + \sigma_{\bf p}) (1 +
	\sigma_{\bf p'}) \enskip S_{\bf p} S_{\bf p'}
	\nonumber \\
	+ \frac{1}{2} \enskip \sum_{{\bf p},{\bf p'}} \enskip V_{\bf pp'}^{c}
	\enskip \sigma_{\bf p} \sigma_{\bf p'}
	\label{H0}
\end{eqnarray}

\noindent where $S_{\bf p} = \pm 1$ is the spin at site $\bf p$. In Eqn.\
(\ref {H0}), $V_{\bf pp'}^{c}$ and $J_{\bf pp'}$ are, respectively,
effective chemical and exchange interactions between sites $\bf p$ and $\bf
p'$. Averaging over the magnetic degrees of freedom, the interacting part
of the alloy Hamiltonian can be written as:

\begin{equation}
	H_2 = \frac{1}{2} \enskip \sum_{{\bf p},{\bf p'}} \enskip \tilde{V}_{\bf
	pp'} \enskip \sigma_{\bf p} \sigma_{\bf p'}
	\label{H2}
\end{equation}

\noindent where the effective pair interaction is:

\begin{equation}
	\tilde{V}_{\bf pp'} = V_{\bf pp'}^{c} \enskip - \enskip \frac{1}{4}
	J_{\bf pp'} <S_{\bf p} S_{\bf p'}>
	\label{V2tilde}
\end{equation}

\noindent We note that in terms of interatomic potentials between different
chemical species, the effective interactions are given by:

\begin{equation}
	\tilde{V}_{\bf pp'} = \frac{1}{4} \enskip [ \tilde{V}_{\bf pp'}^{FeFe} +
	\tilde{V}_{\bf pp'}^{AlAl} - 2 \enskip \tilde{V}_{\bf pp'}^{FeAl}]
	\label{V2}
\end{equation}

Thus, in the present model, the unknown magnetic SRO is included in the
effective pair interactions, which now should depend on temperature.
\cite{pier90} Furthermore, since the product $J_{\bf pp'} <S_{\bf p} S_{\bf
p'}>$ in Eqn.\ (\ref {V2tilde}) is alway positive, the $\tilde{V}_{\bf pp'}$
are smaller than the chemical interactions  $V_{\bf pp'}^{c}$.

\subsection{Cluster Variation Method}

The Hamiltonian in Eqn.\ (\ref{H2}) is solved at finite temperatures using
the Cluster Variation Method. In this approximation, the alloy free energy
is written in terms of the probabilities of finding different atomic
configurations on clusters of lattice sites. A given cluster, consisting of
$n$ lattice points ($ {\bf p}_1, {\bf p}_2, ...,{\bf p}_n$), will be
characterized by a set of indices $\bf n$, describing the geometry and the
orientation of the cluster, and by its location in the lattice, $\vec
r_{\bf n}$. As needed, the three features of the cluster, type and
orientation $\bf n$ and location $\vec r_{\bf p}$, will be collectively
denoted by $\beta = [{\bf n},\vec r_{\bf p}]$. Here we adopt the convention
that the location of the cluster is given by its the center of gravity:

\begin{equation}
	r_{\bf n} = {1 \over n} \enskip \sum_{{\bf p} \in \beta } \enskip
	\vec R_{{\bf p}}
	\label{rn}
\end{equation}¥

For a given cluster $\beta = [{\bf n},\vec r_{\bf n}]$ there are $2^n$
atomic configurations, each one described by the set of occupation numbers
$\vec \sigma_{\beta} = [\sigma_{{\bf p}_1},...\sigma_{{\bf p}_n}]$. The
probability of occurrence of any one of these configurations will be called
$\rho_{\beta}(\vec \sigma_{\beta})$.

With the energy given by Eqn.\ (\ref {H2}), the CVM free energy functional
takes the form: \cite{sanc84}

\begin{eqnarray}
	{\cal F} =  \frac{1}{2} \enskip \sum_{{\bf p},{\bf p'}} \enskip
	\tilde{V}_{\bf pp'} \enskip <\sigma_{\bf p} \sigma_{\bf p'}>
	 + \sum_{\bf p} \mu_{\bf p} <\sigma_{\bf p}>
	\nonumber \\
	- k_B T \enskip \sum_{\beta} \enskip a_{\beta} \enskip
	\sum_{\vec \sigma_{\beta}}
	\enskip
	 \rho_{\beta}(\vec \sigma_{\beta}) \enskip
	 ln \left(\rho_{\beta}(\vec \sigma_{\beta})\right)
	       \label{F}
\end{eqnarray}¥

\noindent where we have included a staggered effective chemical potential
field $\mu_{\bf p}$, which will be used in the next subsection to describe
general concentration fluctuations. The coefficients $ a_{\beta}$ in Eqn.\
(\ref {F}) are central to the CVM. After choosing a set of maximum clusters
$\{ \gamma \}$, which determines the level of approximation, the
coefficients $a_{\beta}$ are calculated from the following recursive
relations: \cite{sanc84}

  \begin{equation}
	  {\sum_{\alpha \supseteq \beta }}^{'} a_{\alpha} =1
	  \label{abeta}
  \end{equation}¥

\noindent where the sum is restricted to clusters $\alpha$ contained in the
set of maximum clusters $\gamma$, i.e. for clusters $\alpha \subseteq \{
\gamma \}$. Furthermore, Eqn.\ (\ref {abeta}) is valid for all clusters
$\beta$ included in the approximation, i.e. $\beta \subseteq \{ \gamma \}$.
The coefficients $ a_{\beta}$ have the full symmetry of the disordered
lattice and, therefore, they are independent of the location and
orientation of the cluster. Furthermore, these coefficients are uniquely
defined once the set of maximum clusters is chosen.

The free energy functional, Eqn.\ (\ref {F}), must be minimized with
respect to the cluster probabilities $\rho_{\beta}(\vec \sigma_{\beta})$
subject to normalization constraints:

\begin{equation}
	\sum_{\vec \sigma_{\alpha}} \enskip \rho_{\alpha}(\vec
	\sigma_{\alpha}) = 1
	\label{norm}
\end{equation}¥

\noindent and, for all clusters $\beta \subseteq \alpha$, subject to the
self consistency requirement:

\begin{equation}
	\rho_{\beta}(\vec \sigma_{\beta}) = {\sum_{\vec \sigma_{\alpha}}}^{'}
	\enskip
	\rho_{\alpha}(\vec \sigma_{\alpha})
	\label{self}
\end{equation}¥

\noindent where the sum is carried out over the occupation numbers
$\sigma_{\bf p} = \pm 1$ for all lattice sites {\bf p} contained in
$\alpha$ but outside cluster $\beta$.

A general description of the cluster probabilities, in which Eqns.\
(\ref{norm}) and (\ref{self}) are strictly obeyed, can be achieved by using
a complete and orthogonal basis set of characteristic functions.
\cite{sanc84,sanc93} These functions, in turn, can be used to describe any
function of configuration, of which the probabilities $\rho_{\beta}(\vec
\sigma_{\beta})$ are a particular case. For binary systems, the
characteristic functions, which we will denote ${\bbox \sigma}_{\beta}$,
are defined for each cluster $\beta$ in the lattice by the product of the
occupation numbers $\sigma_{\bf p}$ over all points $ \bf p$ contained in
$\beta$:\cite{sanc84}

\begin{equation}
		{\bbox \sigma}_{\beta} = \prod_{{\bf p}\in \beta} \enskip
		\sigma_{\bf p}
		\label{charf}
\end{equation}

In this basis, the cluster probabilities take a particularly simple
form:\cite{sanc84}

\begin{equation}
	\rho_{\beta}(\vec \sigma_{\beta}) = {1 \over {2^n}}
	\enskip \left[ 1 +
	\sum_{\alpha \subseteq \beta} {\bbox \sigma}_{\alpha}
	<{\bbox \sigma}_{\alpha}>
	\right]
	\label{rho}
\end{equation}

\noindent where the multisite correlations, $<{\bbox \sigma}_{\alpha}>$,
given by the expectation value of the characteristic functions, form a set
of linearly independent configurational variables. These multisite
correlation can be conveniently used in the free energy minimization:

	\begin{equation}
		{\partial {\cal F} \over \partial <{\bbox \sigma}_{\beta}>} = 0
	 \label{minF}
	\end{equation}

\noindent The minimum of $\cal F$ defines the equilibrium free energy, $F$,
as well as the equilibrium values of the multisite correlation functions,
$<{\bbox \sigma}_{\beta}>$. The latter, of course, have the symmetry of the
equilibrium phase. In the usual case in which the effective chemical
potential field is spatially uniform, the number of correlation functions
$<{\bbox \sigma}_{\beta}>$ is relatively small.   In such cases, the set of
Eqns.\ (\ref {minF}) can be solved efficiently using fast numerical
algorithms based, for example, in the Newton-Raphson method. In the next
subsection, we will study the response of the system to small arbitrary
variations in the effective chemical potential. The analysis leads to the
fluctuation spectrum of the concentration variable $<\sigma_{\bf p}>$,
which is directly related to the observed SRO intensity.

\subsection{Fluctuations}

We note that solution of Eqns.\ (\ref {minF}) yields a set of
correlations or, equivalently, Warren-Cowley SRO parameters for only those
pairs included in the set of maximum clusters $\{ \gamma \}$. Typically,
the Warren-Cowley parameters obtained in this manner are not sufficient to
reproduce the experimental SRO intensity. Thus, the required long-range
pair correlations are calculated using a different approach, based on the
fluctuation spectrum obtained with the CVM free energy, Eqn.\ (\ref {F}).
For the sake of completeness, we summarize here the main results needed for
the calculation of SRO intensities using the CVM, and refer the reader to
the original work for further details. \cite{sanc82} Following Ref.
[\onlinecite{sanc82}], we write the pair correlations as:

\begin{equation}
	<\sigma_{\bf p} \sigma_{{\bf p}'}> - <\sigma_{\bf p}>
	<\sigma_{{\bf p}'}> = - k_B T {\partial^2 F \over {\partial \mu_{\bf p}
	\partial \mu_{{\bf p}'}}}
	\label{sigsigp}
\end{equation}¥

\noindent A Fourier transform of Eqn.\ (\ref {sigsigp}) yields: \cite{sanc82}

\begin{equation}
	<| \sigma (\vec k) |^2> - |<\sigma (\vec k)>|^2  = -k_B T {\partial^2 F
	\over {\partial \mu (\vec k) \partial \mu (- \vec k)}}
	\label{sigsigk}
\end{equation}¥

with:

\begin{equation}
	\sigma (\vec k) = \sum_{\bf p} \enskip \sigma_{\bf p} \enskip  e^{-i \vec k
	\cdot \vec R_{\bf p}}
	\label{sigk}
\end{equation}¥

and

\begin{equation}
	{\partial \over \partial \mu (\vec k)} =  \sum_{\bf p} \enskip e^{-i \vec k
		\cdot \vec R_{\bf p}} {\partial \over \partial \mu_{\bf p}}
		\label{dmu}
\end{equation}¥

Actual calculation of the right hand side of Eqn.\ (\ref {sigsigk})
requires the Fourier transform of the matrix of second derivatives of the
free energy with respect to the correlation functions:

\begin{equation}
	F_{{\bf n},{\bf n'}}(\vec k) = {1 \over N} \enskip
	\sum_{\vec r_{\bf n}, \vec r_{\bf n'}} \enskip
	F_{{\bf n},{\bf n'}}(\vec r_{\bf n} - \vec r_{\bf n'})
	\enskip
	 e^{-i \vec k \cdot (\vec r_{\bf n} - \vec r_{\bf n'})}
		\label{Fnnk}
\end{equation}¥

\noindent where the sum is over all the locations $\vec r_{\bf n}$ and
$\vec r_{\bf n'}$ in the lattice of clusters of type $\bf n$ and $\bf n'$
and where $F_{{\bf n},{\bf n'}}(\vec r_{\bf n} - \vec r_{\bf n'})$ is given
by:

	\begin{equation}
		F_{{\bf n},{\bf n'}}(\vec r_{\bf n} - \vec r_{\bf n'}) =
		{\partial^2 {\cal F}\over {\partial <{\bbox \sigma}_{\beta}>
		\partial <{\bbox \sigma}_{\beta '}>}}
		\label{FnnR}
	\end{equation}

\noindent with $\beta =[ {\bf n}, \vec r_{\bf n}] $ and $\beta '=[ {\bf n '},
\vec
r_{\bf n'}] $.

Since the effective chemical potential $\mu (\vec k)$ is the conjugate
thermodynamic variable of the point correlation function $<\sigma (\vec
k)>$, the second derivatives in Eqn.\ (\ref{sigsigk}) are given by:
\cite{sanc82}

\begin{equation}
	-{\partial^2 F \over {\partial \mu (\vec k) \partial \mu (- \vec k)}}
	= N F_{1,1}^{-1}(\vec k)
	\label{F11}
\end{equation}¥

\noindent where $F_{1,1}^{-1}(\vec k)$ is the diagonal element
corresponding to the point correlation function (${\bf n} = {\bf n'} = 1$)
of the inverse of the matrix $ F_{{\bf n},{\bf n'}}(\vec k)$.

Combining Eqns.\ (\ref{alphap}), (\ref{sigsigk}) an (\ref{F11}), we obtain
the desired result to be used in the fitting of the SRO intensity:

\begin{equation}
	\alpha (\vec k) = {N k_B T
	F_{1,1}^{-1}(\vec k) \over  {1 - < \sigma_{\bf o}>^2} }
	\label{alphafit}
\end{equation}

The procedure leading to Eqn.\ (\ref {alphafit}) is equivalent to expanding
the free energy around equilibrium up to second order in the correlation
functions, and describing the fluctuation spectrum in the Gaussian
approximation. As mentioned, Eqn.\ (\ref {alphafit}) yields a significantly
improved description of the SRO intensity over the Krivoglaz-Clapp-Moss
formula. We note that although accuracy is generally gained as the cluster
size increases, the size of the matrix of second derivatives $ F_{{\bf
n},{\bf n'}}(\vec k)$ can rapidly grow beyond practical limits. In the next
section, we will describe results for a CVM approximation that include
clusters of up to nine points, which yields a complex second derivative
matrix of size $1088 \times 1088$.

\subsection{Inverse k-Space Method}¥

The simplest procedure to obtain effective pair interactions from the
measured SRO intensity consists in calculating the Warren-Cowley SRO
parameters $\alpha_{\bf p}$ from the measured $\alpha (\vec k)$, and then
proceed to solve for the interactions using Eqns.\ (\ref {minF}) with the
proper constraints on the pair probabilities. As mentioned, this real space
inverse method has been applied to several cases using both the CVM and
Monte Carlo simulations. Here we propose an alternative approach that does
not require the computation of the real space Warren-Cowley SRO parameters
but, instead, obtains the pair interactions by fitting $\alpha (\vec k)$,
calculated using Eqn.\ (\ref {alphafit}), to those obtained experimentally.
The procedure consists in minimizing the following sum of errors squared:

\begin{equation}
	\chi ^2 = {1 \over M} \enskip \sum_{i=1}^M \enskip
	\left[{  {\alpha_{m}(\vec k_i) - \alpha_{c}(\vec k_i)} \over
	{\Delta \alpha(\vec k_i)} }\right]^2
	\label{chi2}
\end{equation}

\noindent with $M$ the number of measurements, $\alpha_{m}(\vec k_i)$ and
$\alpha_{c}(\vec k_i)$ the measured and calculated SRO intensities,
respectively, and where $\Delta \alpha(\vec k_i)$ is the experimental
errors determined as described in Section II.

In order to estimate the errors in the effective interactions $\tilde
{V}_n$ obtained from the least square fitting, we consider the probability
distribution for the SRO intensity to be of the form:

\begin{equation}
	P(\{ \alpha (\vec k) \}) \propto \prod_{i=1}^{M} \enskip
	exp \left[- { \delta \alpha (\vec k_i)^2 \over
	\Delta {\alpha (\vec k_i)}^2 }\right]
	\label{Palpha}
\end{equation}¥

\noindent with $\delta \alpha (\vec k_i) = \alpha (\vec k_i) - <\alpha
(\vec k_i)>$.

 The corresponding probability distribution for the effective
interactions, up to second order in $\tilde V_n$, is then given by:

\begin{equation}
	P( \{ {\tilde V}_n \}) \propto \prod_{n} \enskip
	exp \left[-{
	\sum_{m,n} A_{m,n}
	{\delta {\tilde V}_{m} \over \Delta {\tilde V}_{m}}
    {\delta {\tilde V}_{n} \over \Delta {\tilde V}_{n}}
    }\right]
	\label{PVn}
\end{equation}

\noindent with $\delta {\tilde V}_{m} = {\tilde V}_{m} - <{\tilde
V}_{m}>$ and  $A_{m,n}$ given by:

\begin{equation}
	A_{n,m} =  \sum_{i=1}^{M} \enskip {1 \over {\Delta \alpha (\vec k_i)}^2}
	\enskip
	{ \partial \alpha (\vec k_i) \over \partial {\tilde V}_n }
    { \partial \alpha (\vec k_i) \over \partial {\tilde V}_m }
	\label{Amn}
\end{equation}

\noindent Thus, an estimate of the error in the effective interactions can
be obtained from the eigenvalues, $\lambda_n$, and the matrix of
eigenvectors, $\vec \nu_{n,m}$, of $A_{m,n}$:

\begin{equation}
	\Delta {\tilde V}_n = \chi \enskip \sum_{m} \enskip {| \nu_{n,m} | \over
	\sqrt{\lambda _n}}
	\label{DeltaV}
\end{equation}¥

\noindent where we have included the factor $\chi$, which is typically
larger than one, in order to take into account the quality of the fit in
the error of the effective interactions.

\section{Results}

The methods described in the previous section were applied to the three
different approximations of the CVM, which will be referred to as the
tetrahedron (T), the cube-octahedron (C-O) and the cube-rhombohedron-octahedron
(C-R-O) approximations. The maximum cluster for each of these
approximations are shown in Fig.\ 2. In the T approximation, only first and
second neighbors are involved. The C-O approximation, which has been used
previously in the context of the real space inverse CVM,
\cite{pier90,pier91} extends the range up to fifth neighbors. In this
approximation, however, the fourth neighbor pairs are excluded since they
do not belong to the maximum clusters (the body-centered-cube and the
octahedron). This difficulty is resolved by the C-R-O approximation since
the additional rhombohedron cluster includes fifth neighbor pairs.

\subsection{Real Space Method}

First, we apply the standard inverse CVM. For comparison purposes we have
carried out the calculations in all three approximation, although the T
approximation is not expected to be very reliable in this case since only
two Warren-Cowley parameters are used in the fit. The results are shown in
Table\ I where the first column lists the Warren-Cowley SRO parameters
extracted from the experimental data for the first 5 neighbor shells. We
recall that 8 shells of $\alpha_i$'s have been used in the experimental
analysis of the data. \cite{pier90,pier91} We also note that in the C-R-O
approximation all the $\alpha_i$'s listed in Table\ I are reproduced
exactly, whereas in the C-O approximation there is no information on
$\alpha_4$ and in the T approximation only the first 2 $\alpha_i$'s are
reproduced.

The values of ${\tilde V}_n$ are comparable in the three approximations,
with the notable exception of ${\tilde V}_2$ that is three to four times
smaller in the C-R-O than in the other two approximations. A direct test of
the quality of the inverse CVM results can be obtained by calculating
$\alpha (\vec k)$, as described in Section III.C, using the interactions of
Table\ I. The results of these calculations for the three
approximation are compared to the experimental data in Fig.\ 3.

Overall, the quality of the fit for all three approximations is not very
satisfactory. The T approximation (dotted line in Fig.\ 3), with only two
effective interactions, gives a very poor agreement near the maximum at
$(0.5,0.5,0.5)$. On the other hand, the C-O approximation (dashed line in
Fig.\ 3) reproduces this maximum quite well, although it fails to reproduce
the increase in intensity predicted by the experimental $\alpha_n$'s (see
Fig.\ 1). As mentioned, the poor fit near the origin may not be
particularly important since, based on the k-space data, the existence of a
maximum near the origin is questionable. Nevertheless, this failure of the
C-O approximation to reproduce the diffuse intensity maximum illustrates
two problematic aspects of the real space inverse CVM: i) The fact that the
cut-off in SRO may introduce spurious effects in the SRO intensity and, ii)
The fact that fitting just a few of the Warren-Cowley SRO parameters is not
enough to reproduce all the features of the SRO intensity.

The inverse CVM in the C-R-O approximation (full line in Fig.\ 3) works
very well around $(0.5,0.5,0.5)$, reproduces the diffuse intensity maximum
near the origin, but it gives poor agreement around (001).

Finally, in order to illustrate the problems that may arise from using the
C-O approximations, in which fourth neighbor pairs are not considered (i.e.
${\tilde V}_4 =0$), we have calculated the Warren-Cowley SRO parameters
using the C-R-O approximation with the interactions obtained in the C-O
approximation. The results are shown in the last column of Table\ I. We see
that the $\alpha_4$ predicted by the interactions of the C-O approximation
differs in sign from the experimental one, given in the second column in
Table\ I.

\subsection{Reciprocal Space Method}

The difficulties encountered in the application of the inverse CVM to the
$Fe-Al$ data are likely due to the uncertainty and errors in the
experimental determination of the $\alpha_n$'s. Although this difficulty
can in principle be resolved by measuring more points on several reciprocal
space planes, a more elegant solution is to fit the SRO intensity directly
in k-space. The resulting pair interactions at $1273K$, together with the
predicted $\alpha_n$'s, are shown in Table\ II. We see from the table that
the interactions, particularly for the nearest- and next-nearest-neighbor
pairs, are significantly different from those obtained using the real space
method. Another noticeable difference is that the interactions shown in
Table\ II change very little as we go from one approximation to the next.
In Figure\ 4 we show a comparison between the results of the real space
(open symbols) and reciprocal space (full symbols) methods for the T
(triangle), C-O (circles) and C-R-O (squares) approximations. The lines
connecting the ${\tilde V}_n$'s obtained in the C-R-O approximation are
drawn as an aid to the eye only.

The SRO intensity corresponding to the interactions of Table\ II are shown
in Fig.\ 5. We see that the overall fit has improved considerably relative
to that obtained with the real space method. Furthermore, the three
approximations give essentially the same intensity. This is understood from
the fact that the dominant interactions are the first and second neighbors
(see Table\ II). Thus, the T approximation is already sufficient to capture
the main features of the diffuse intensity. The increase in intensity near
the origin predicted by the experimental $\alpha_n$'s is not present in any
of the three approximations.

Since the dominant effective interactions are between first and second
neighbors, the SRO intensity for the other temperatures investigated here
were analyzed using the T approximation. The results for ${\tilde V}_1$ and
${\tilde V}_2$ as a function of temperature are shown in Fig.\ 6. There is
a gradual increase in the effective interactions with temperature which is
expected from the decrease in magnetic SRO (see Eqn.\ (\ref{V2tilde})).

\section{Conclusions}

The determination of effective interactions from measured SRO intensities
requires considerable manipulation of the experimental data. In this
process, the guidance of reliable theoretical models is invaluable. In
particular, we have shown that the ability to calculate accurately the
diffuse intensity in reciprocal space allows the determination of the
effective interactions without the intermediate step of obtaining the
Warren-Cowley SRO parameters in real space. Consequently, the method
proposed here avoids the need to introduce a real space cut-off in SRO.

The study of fluctuations in reciprocal space by means of the CVM allows
the description of statistical correlations beyond the range of effective
interactions. This establishes a fundamental difference with the KCM
formula, derived from the Bragg-Williams approximation, in which the
fluctuation spectrum is determined entirely by the Fourier transform of the
interaction potentials. Furthermore, these statistical correlations, which
extend beyond the range of the effective interactions, are describe well
using relatively small clusters. For example, for the $Fe-Al$ studied here,
and due to the fact that first and second pair interactions are dominant,
we find that the Tetrahedron approximation suffices for an accurate
description of the experimental data.

As an illustration of the advantages of the k-space versus the real space
methods, we note that the k-space method in three different approximations
produces essentially the same results in terms of the effective potentials
and predicted SRO intensities. In contrast, the results of the real space
CVM are, for the same data, significantly different to each other and to
the results of the k-space analysis. Particularly revealing is the fact
that increasing the number of Warren-Cowley SRO parameters used as input in
the real space inverse CVM does not necessarily produce better results. For
example the inaccuracies of the T approximation which, incidentally, works
well in the k-space analysis, can in principle be explained by the fact
that only $\alpha_1$ and $\alpha_2$ are considered. Increasing the
$\alpha_n$'s to five in the C-R-O approximation improves the results in
some regions in reciprocal space but also introduce large errors in other
regions and, possibly, spurious effects such as the increase in intensity
near the origin. In fact, for the case studied here, the C-O approximation,
that neglects the fourth neighbor interactions altogether, appears to
produce the more sensible results in terms of the predicted intensities.
However, a closer look shows that the interactions obtained in the C-O
approximation give the wrong sign of the fourth neighbor Warren-Cowley
parameter.

Finally, we point out that the k-space CVM analysis of the SRO intensity calls
for a different experimental approach. Whereas the real space methods
require measurements of $\alpha(\vec k)$ in many points in the irreducible
Brillouin zone in order to perform the Fourier transformation, only a few
points are actually needed when using the k-space CVM analysis. Thus, a
more efficient strategy to obtain effective interactions is to measure the
diffuse scattering at fewer points in reciprocal space but with higher
accuracy.

\section{Acknowledgments} The work at The University of Texas at Austin was
supported by the National Science Foundation under Grants No. DMR-91-14646
and No. INT-91- 14645.

\begin{figure}
\epsfbox[60 200 550 600]{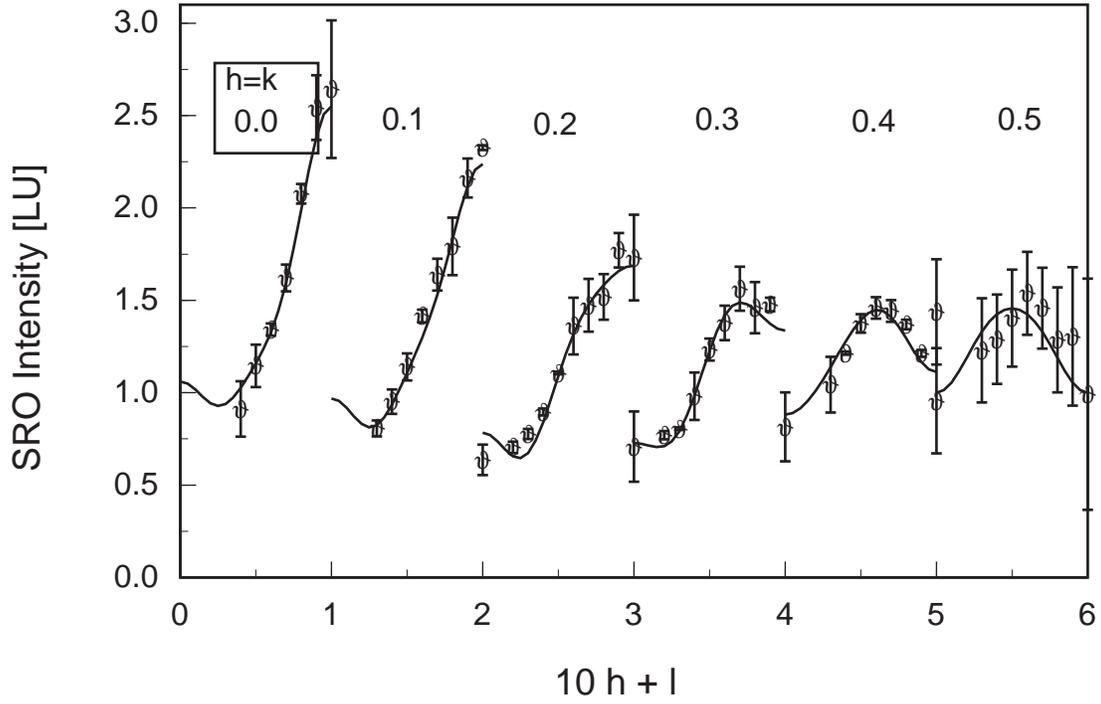}
\caption{%
Experimental short-range order diffuse intensity (symbols) and the fit
(solid line) obtained using 8 shells of Warren-Cowley short-range
parameters $\alpha_n$.}
\end{figure}

\begin{figure}
\epsfbox[60 200 550 600]{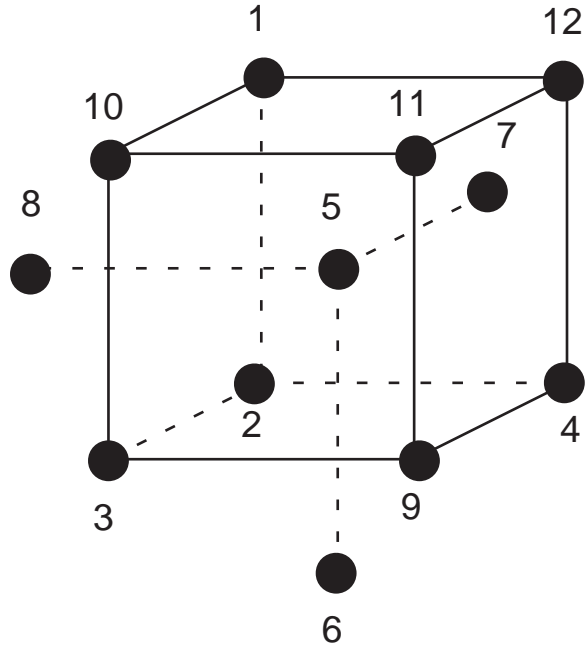}
\caption{%
Clusters used in the three different approximations of the Cluster
Variation Method: Tetrahedron (T) approximation (points 1-2-5-7);
Cube-Octahedron (C-O) approximation combining the body-centered cube
(points 1-2-3-4-5-9-10-11-12) and octahedron (points 2-3-4-5-6-9) clusters;
and the Cube-Rhombohedron-Octahedron (C-R-O) approximation combining, in
addition to the two clusters in the Cube-Octahedron approximation, the
rhombohedron labeled 1 through 8.}
\end{figure}

\begin{figure}
\epsfbox[60 200 550 600]{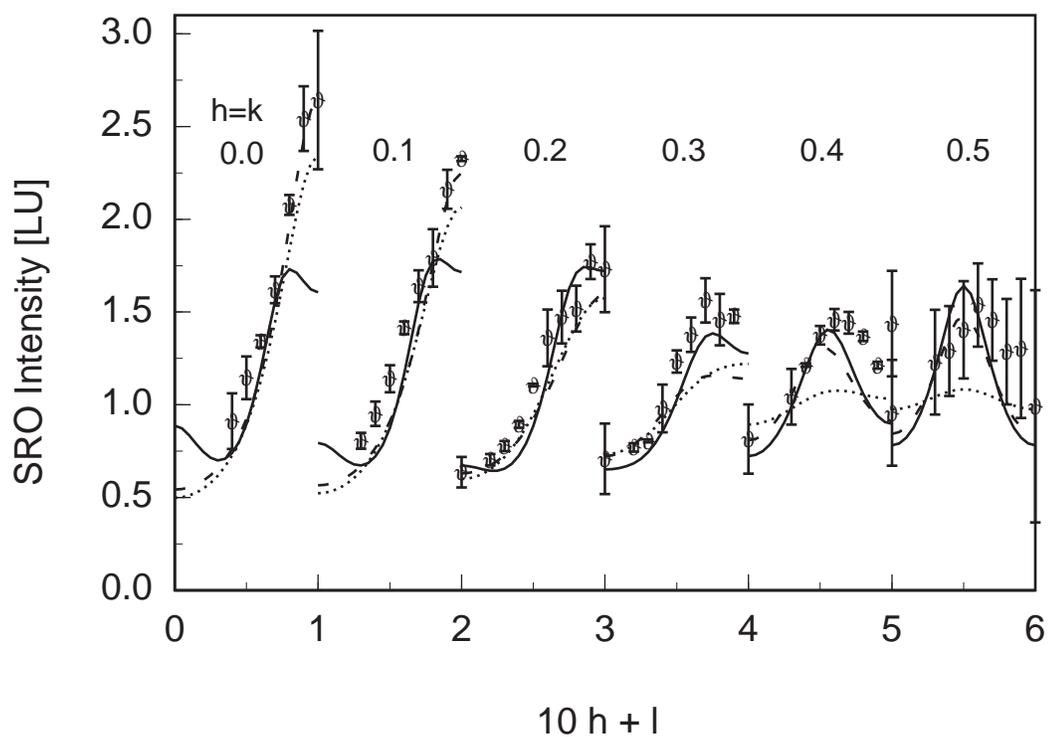}
\caption{%
Short range order diffuse intensity obtained using the real space inverse
cluster variation method in three approximations: Tetrahedron (dotted
line); Cube-Octahedron (dashed line) and Cube-Rhombohedron-Octahedron (full
line). The symbols are experimental points.}
\end{figure}

\begin{figure}
\epsfbox[60 200 550 600]{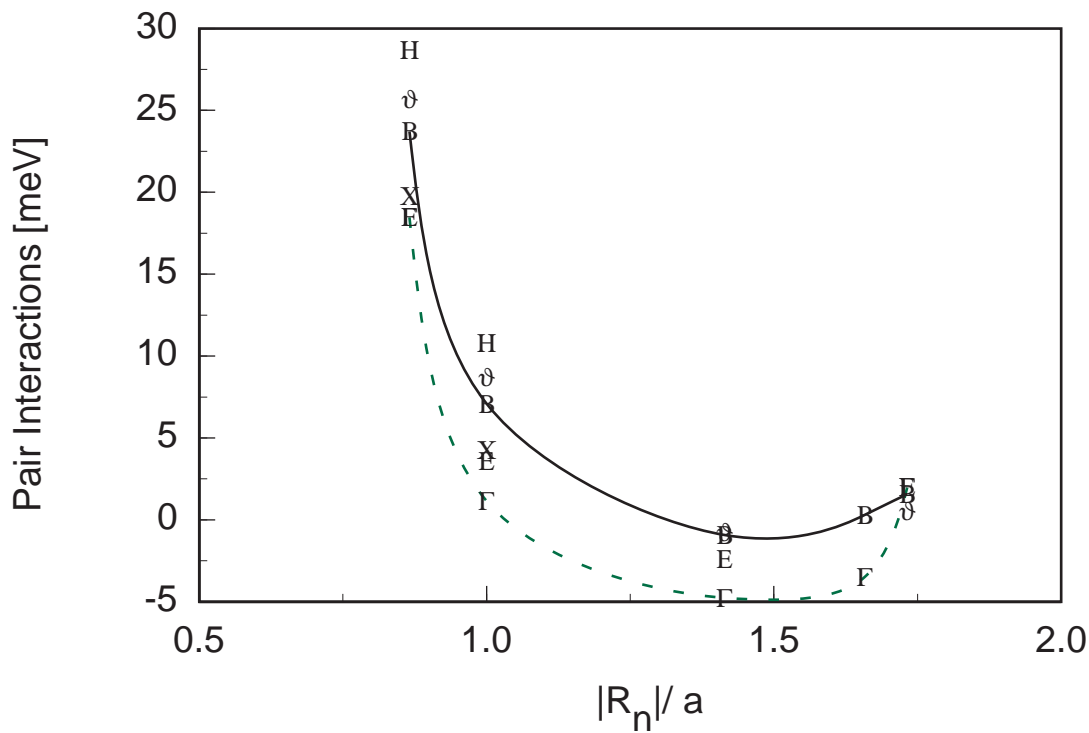}
\caption{%
Effective pair interactions as function of distance, in units of the
lattice constant $a$, obtained using the real space (open symbols) and the
k-space (full symbols) methods described in the text. For each method, the
cluster variation approximations used were the Tetrahedron (triangles),
Cube-Octahedron (circles) and Cube-Rhombohedron-Octahedron (squares).}
\end{figure}

\epsfbox[60 200 550 600]{Fig5}
\begin{figure}
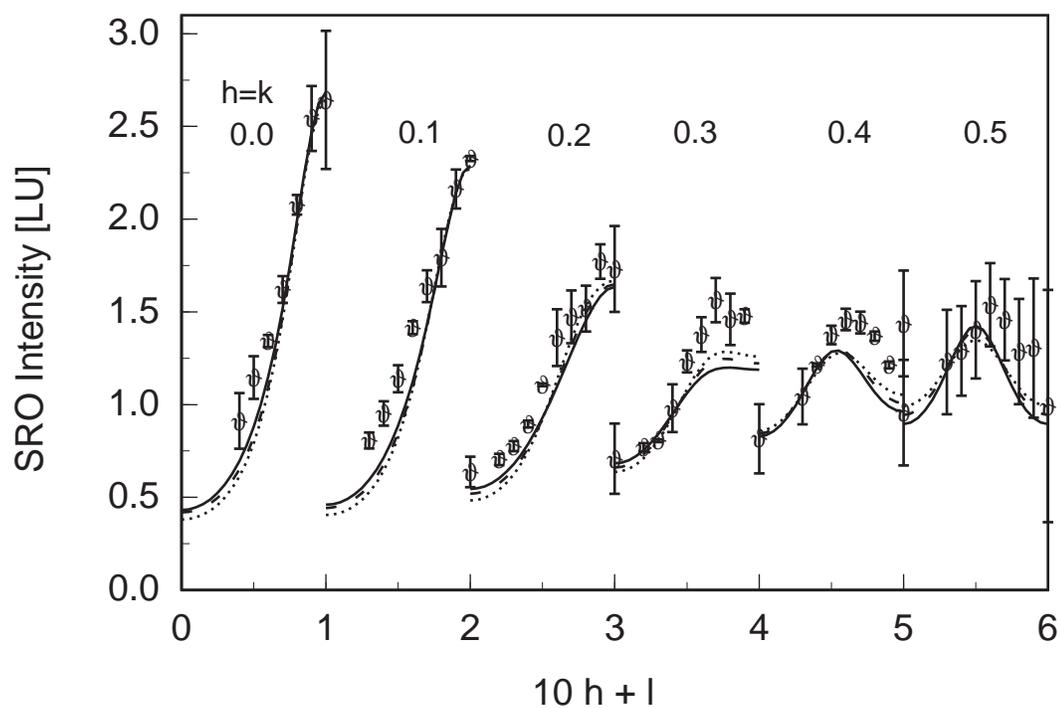

\caption{%
Short range order diffuse intensity obtained using the reciprocal space
cluster variation method described in the text in three approximations:
Tetrahedron (dotted line); Cube-Octahedron (dashed line) and
Cube-Rhombohedron-Octahedron (full line). The symbols are experimental points.}
\end{figure}

\begin{figure}
\epsfbox[60 200 550 600]{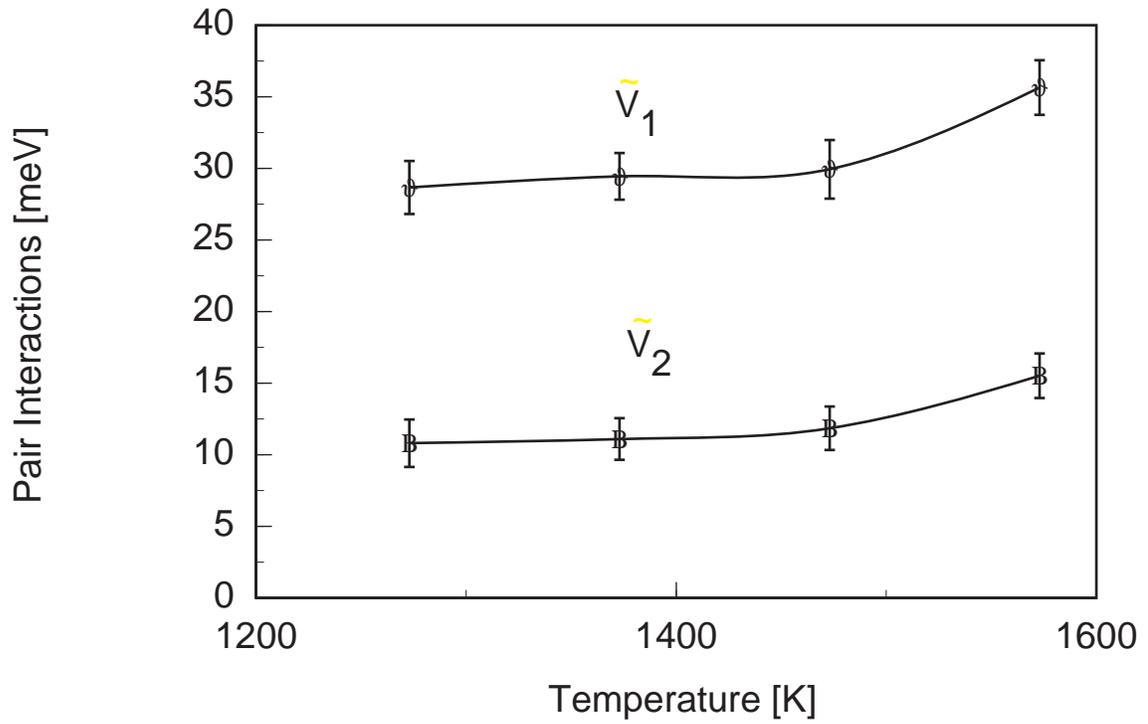}
\caption{%
Temperature variation of the effective pair interactions for first
(${\tilde V}_1$) and  second (${\tilde V}_2$) neighbors obtained using
the reciprocal space method in the Tetrahedron approximation.}
\end{figure}

\begin{table}
\caption%
{Pair interactions $\tilde V_n$ for ${\bf Fe}19.5 at\% Al$ at $1273 K$
obtained using real space inversion for three approximations of the cluster
variation method. Errors in the estimated pair interactions are shown in
parenthesis. Experimental Warren-Cowley short range order parameters are
given in the second column. Those in the last column were calculated in the
C-R-O approximation with the interactions of the C-O approximation.}
\begin{tabular}{ccdddc}
&\multicolumn{1}{c}{$Exp.$}&\multicolumn{1}{c}{$T$}&
\multicolumn{1}{c}{$C-O$}
&\multicolumn{2}{c}{$C-R-O$}\\
\tableline
n&$\alpha_{i}$&$\tilde V_n (meV)$&$\tilde V_n (meV)$&$\tilde V_n (meV)$&
$\alpha_{i}$\\
\tableline
 1 & -0.0943 & 19.7(0.2)  &   18.5(0.3)  &   18.5(0.2) & -0.0944 \\
 2 &  0.0114 &  4.2(0.2)  &    3.6(0.2)  &    1.1(0.2) &  0.0113 \\
 3 &  0.0345 &            & $-$2.4(0.1)  & $-$4.8(0.1) &  0.0351 \\
 4 &  0.0097 &            &              & $-$3.5(0.1) & -0.0091 \\
 5 & -0.0020 &            &    2.0(0.1)  &    2.0(0.1) & -0.0019 \\
\end{tabular}
\label{table1}
\end{table}

\begin{table}
\caption
{Pair interactions $\tilde V_n$ and the corresponding Warren-Cowley
short-range order parameters for ${\bf Fe}19.5 at\% Al$ at $1273 K$ obtained
by fitting the SRO intensity in k-space using three approximations of the
cluster variation method. Errors in the estimated pair interactions are
shown in parenthesis.}
\begin{tabular}{cdcdcdc}
&\multicolumn{2}{c}{$T$}&\multicolumn{2}{c}{$C-O$}
&\multicolumn{2}{c}{$C-R-O$}\\
\tableline
n&$\tilde V_n (meV)$&$\alpha_{n}$&$\tilde V_n (meV)$&$\alpha_{n}$&
$\tilde V_n (meV)$&$\alpha_{n}$ \\
\tableline
 1 & 28.7(1.9) & -0.1158 &   25.7(4.7) & -0.1107  &   23.7(1.9) & -0.1067 \\
 2 & 10.8(1.7) & -0.0044 &    8.7(3.7) & -0.0005  &    7.1(0.4) &  0.0047 \\
 3 &           &         & $-$0.7(1.2) &  0.0326  & $-$0.9(0.1) &  0.0332 \\
 4 &           &         &             &          &    0.3(0.1) & -0.0080 \\
 5 &           &         &    0.6(1.8) &  0.0076  &    1.6(0.1) &  0.0016 \\
\end{tabular}
\label{table2}
\end{table}
\end{document}